\documentclass[aps,prl,twocolumn,10pt,showpacs,superscriptaddress]{revtex4-1}
\usepackage[T1]{fontenc} \usepackage[english]{babel}
\usepackage{amsmath,amssymb,amsbsy,graphicx,braket,mathtools,wrapfig,cleveref,color,epsfig,import,bm,silence}
\renewcommand{\v}[1]{\ensuremath{\mathbf{#1}}} 
\newcommand{\vg}[1]{\ensuremath{\boldsymbol{#1}}}
\newcommand{\D}{^{\dag}}
\newcommand{\p}{^{\prime}}

\newcommand{\conj}{^*}
\newcommand{\MD}{^{\mathstrut}}
\newcommand*\diff{\mathop{}\!\mathrm{d}}

\providecommand*{\iu}{\ensuremath{\mathrm{i}}}
\DeclareMathOperator{\sign}{sign}

\DeclarePairedDelimiter\abs{\lvert}{\rvert}
\newcommand{\vssp}{\vspace{0.1cm}}

\begin{document}
\title{Quantum Quenches in Chern Insulators} \author{M. D. Caio}
\affiliation{Department of Physics, King's College London, Strand,
  London WC2R 2LS, United Kingdom} \author{N. R. Cooper}
\affiliation{T.C.M. Group, Cavendish Laboratory, J.J. Thomson Avenue,
  Cambridge CB3 0HE, United Kingdom} \author{M. J. Bhaseen}
\affiliation{Department of Physics, King's College London, Strand,
  London WC2R 2LS, United Kingdom}


\begin{abstract}
We explore the non-equilibrium response of Chern insulators.
Focusing on the Haldane model, we study the dynamics induced by 
quantum quenches between topological and non-topological phases. 
A notable feature is that the Chern number, calculated for an
infinite system, is unchanged under the dynamics following such a quench. 
However, in finite geometries, the initial and final Hamiltonians 
are distinguished by the presence or absence of edge modes. 
We study the edge excitations and describe their impact on 
the experimentally-observable edge currents and magnetization. 
We show that, following a quantum quench, the edge currents relax 
towards new equilibrium values, and that there is light-cone
spreading of the currents into the interior of the sample.
\end{abstract}

\maketitle

Topological phases of matter display many striking features, ranging
from the precise quantization of macroscopic properties, to the
emergence of fractional excitations and gapless edge states.  An
important class of topological systems is provided by the so-called
Chern insulators realized in two-dimensional settings
\cite{Thouless1982}. A famous example is the Haldane model
\cite{Haldane1988}, which describes spinless fermions hopping on a
honeycomb lattice. The Haldane model exhibits both topological and
non-topological phases, and its behavior is closely related to the
integer quantum Hall effect. Recent advances using ultra cold atoms
\cite{Goldman2009,Alba2011,Aidelsburger2013,Goldman2013,Wright2013}
have led to the experimental realization of the Haldane model
\cite{Esslinger2014}. Proposals also exist for realizing other states
of topological matter using cold atoms \cite{Setiawan2015}.

A fundamental characteristic of topological systems is their
robustness to local perturbations, making them ideal candidates for
applications in metrology and quantum computation. However, much less
is known about their dynamical response to global perturbations and
time-dependent driving. This issue is of relevance in a variety of
contexts, ranging from the time-evolution and controlled manipulation
of prepared topological states, to the dynamics of topological systems
coupled to their environment. Understanding the impact of topology on
the out of equilibrium response is crucial for further developments,
and is the motivation for this present work. For recent progress in
this direction see
Refs.~\cite{Dora2010,Uehlinger2013,Barnett2013,Patel2013,Slingerland2014,Dutta2010,Hauke2014,Stojchevska2014}.

In this manuscript we investigate the non-equilibrium dynamics of the
paradigmatic Haldane model. In particular, we consider quantum
quenches and sweeps between topological and non-topological
phases. Key questions that we will address include: What happens to
the topological properties on transiting between different phases?
What happens to the edge excitations following a quantum quench?
How do the topological characteristics influence
the non-equilibrium dynamics?

\vssp
\begin{figure}[ht]
  \centering  
   \includegraphics{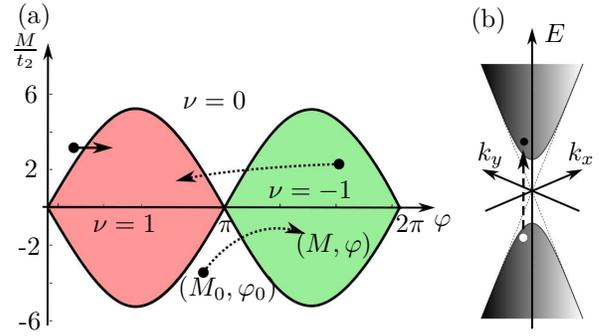}
   \caption{(a) Phase diagram of the Haldane model obtained from the
       low-energy Dirac fermion representation, showing topological
       ($\nu=\pm 1$) and non-topological phases
       ($\nu=0$)~\cite{Haldane1988}. We consider quantum quenches and
       sweeps between different regions of the phase diagram, as
       illustrated by the arrows. (b) The low-energy spectrum of the
       Haldane model is described by excitations around two Dirac
       points.  After a quench, carriers in the lower band are excited
       to the upper band.}\label{fig:Hal_Ch}
 \end{figure}
\emph{Model.}--- The Haldane model describes spinless fermions hopping
on a honeycomb lattice with both nearest and next nearest neighbor
hopping parameters. The Hamiltonian is given by \cite{Haldane1988}
\begin{multline}
 \hat H=t_1\sum_{\langle i,j\rangle}\left(\hat c\D_i \hat c\MD_j+\mbox{h.c.}\right)\\
   + t_2 \sum_{\langle\!\langle i,j\rangle\!\rangle}\left(e^{\iu \varphi_{ij}} \hat c\D_i \hat c\MD_j+\mbox{h.c.}\right)\\
   +M \sum_{i\in A} \hat n_i -M \sum_{i\in B} \hat n_i,\label{eq:HM_realspace}
\end{multline}
where the fermionic operators obey the anticommutation relations
$\{\hat c_j,\hat c_j^\dagger\}=\delta_{ij}$ and $\hat n_i\equiv \hat
c_i^\dagger{\hat c_i}$. Here, $\langle i,j\rangle$ and
$\langle\!\langle i,j\rangle\!\rangle$ indicate the summation over the
nearest and next to nearest neighbor sites respectively, and $A$ and
$B$ label the two sub-lattices. The phase factor
$\varphi_{ij}=\pm\varphi$ is introduced in order to break
time-reversal symmetry and is positive for anticlockwise next to
nearest neighbor hopping. The energy off-set $\pm M$ breaks spatial
inversion symmetry. The phase diagram of the Haldane model is shown in
Fig.~\ref{fig:Hal_Ch} (a); following Ref.~\cite{Haldane1988} we assume
that $|t_2/t_1|\le 1/3$ so that the bands may touch, but not overlap.

For $t_2,M\ll t_1$, the Hamiltonian (\ref{eq:HM_realspace})
has a linear dispersion near the six corners of the hexagonal 
Brillouin zone, but only two of these are inequivalent. 
As a result, close to half-filling, the low-energy description 
is given by the sum of two Dirac Hamiltonians
\begin{equation}
\hat H_\alpha=\begin{pmatrix}
    m_\alpha c^2&-c\,k\, e^{\iu \alpha \theta}\\-c\,k\, e^{-\iu \alpha\theta} & -m_\alpha c^2
   \end{pmatrix},\label{eq:dirac_matrix}
\end{equation}
where $\alpha=\pm 1$ label the Dirac points.  Here, $c=3t_1/2\hbar$ is
the effective speed of light, $k \exp(\iu \theta)$ parameterizes the
2D momentum $(k_x,k_y)$ 
and $m_\alpha =(M-3\sqrt{3}\alpha t_2 \sin \varphi)/c^2$ is the effective
mass \cite{Haldane1988}. The topological phases have a non-vanishing
Chern number $\nu$
\cite{Chern1946,Thouless1982,Berry1984,Haldane1988}. For a state
$|\psi\rangle$ this is defined by the integral of the Berry curvature
over the 2D Brillouin zone
\begin{equation}
\nu=\frac{1}{2\pi}\int d^2k \,\Omega,
\label{eq:chernomega}
\end{equation}
where
$\Omega=\partial_{k_x}A_{k_y}-\partial_{k_y}A_{k_x}$ and $A_{k_\mu} =\iu
\Braket{\psi|\partial_{k_\mu}|\psi}$ is the Berry connection. For the
ground state of the Haldane model $\nu\in \pm 1,0$.  This may be
decomposed into contributions from the two Dirac points as
$\nu=\nu_++\nu_-$, where $\nu_\alpha=-\tfrac{\alpha}{2}\,{\rm
  sign}(m_\alpha)\in \pm 1/2$. The boundaries of the topological
phases correspond to the locations where $m_\pm$ changes sign. They 
are thus given by $M/t_2=\pm 3\sqrt{3} \sin \varphi$, and are
independent of $t_1$; see Fig.~\ref{fig:Hal_Ch}. 

\vssp
{\em Quantum Quenches.}--- In order to gain insight into the
non-equilibrium dynamics of the Haldane model, we consider quantum
quenches between different points $(M,\varphi)$ on the phase diagram
shown in Fig.~\ref{fig:Hal_Ch}, for fixed values of $t_1$ and
$t_2$. At time $t=0$, we prepare our system in the ground state with
parameters $(M_0,\varphi_0)$. At half-filling our initial state
fills the lower band. We then abruptly change the
parameters of $\hat H$ to $(M,\varphi)$, and allow the system to
evolve unitarily under the action of this new Hamiltonian. In general,
this will lead to a non-trivial occupation of both the lower and the
upper bands.

\begin{figure}
  \centering  
   \includegraphics{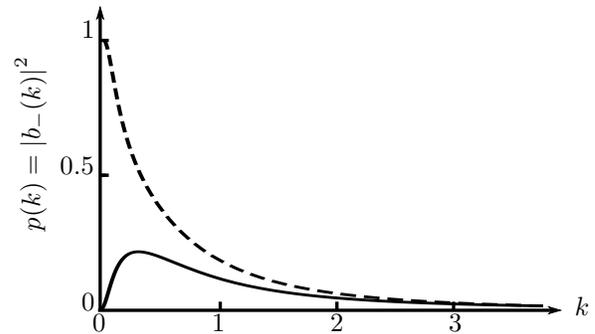}
     \caption{Probability of occupying the upper band for a single
       Dirac point ($\alpha=-1$) with $c=1$, following a quench of
       $m_\alpha$.  Sign-preserving quench, $m_-=-1\rightarrow
       m_-^\prime=-0.1$ (solid line) and a sign-changing quench,
       $m_-=-1\rightarrow m_-^\prime=0.1$ (dashed line). In both
       cases, $b_-(\infty)=0$, corresponding to the time-independence
       of $\nu_-(t)$ using Eq.~(\ref{eq:chern_number}).  The
       sign-changing quench yields $|b_-(0)|^2=1$, but the
       contribution to $\nu_-(t)$ in Eq.~(\ref{eq:chern_number}) is
       compensated by the change in sign of $m_-$. As a result,
       $\nu_-$ is unchanged from its initial value.}
\label{fig:bsquare}
 \end{figure}

We begin by examining the non-equilibrium response of
the effective Dirac Hamiltonian ${\hat H}={\hat H}_++{\hat H}_-$.
Since $\nu=-\tfrac{1}{2}\left[{\rm sign}(m_+)-{\rm sign}(m_-)\right]$,
quenching between different phases corresponds to changing the sign of
one or both of the masses $m_\alpha$. For a given Dirac
point, such changes will lead to a re-distribution of
carriers between the two bands. For a $\theta$ independent
superposition, $\Ket{\psi_\alpha(k)}=a_\alpha(k)e^{-\iu
  E_\alpha^{l}(k) t}\Ket{l_\alpha(k)}+b_\alpha(k)e^{-\iu
  E_\alpha^{u}(k) t}\Ket{u_\alpha(k)}$, the Chern number is formally
given by
\begin{multline}
\nu_\alpha(t)=-\alpha\sign m_\alpha \left(\frac{1}{2}-\abs{b_\alpha(0)}^2
\right) \\- \abs{b_\alpha(\infty)}\abs{a_\alpha(\infty)} \cos [
  (E_\alpha^{u}(\infty)-E_\alpha^{l}(\infty))t+\delta]. \label{eq:chern_number}
\end{multline}
Here, $a_\alpha(k)$ and $b_\alpha(k)$ are complex $c$-numbers,
$\delta={\rm arg}(a_\alpha(\infty))-{\rm arg}(b_\alpha(\infty))$, and
$E_\alpha^{l,u}(k)$ are the energies in the lower and upper bands.
In general, $\nu_\alpha(t)$ is time-dependent,
and differs from its ground state values $\pm 1/2$. However, the
time-dependence only enters via the superposition coefficients
evaluated at $k=\infty$.  An explicit computation shows that
$b_\alpha(\infty)=0$, following a quantum quench; see
Fig.~\ref{fig:bsquare} and the Supplemental Material.  In addition,
$b_\alpha(0)=0,\pm 1$, so the potential modification of $\nu_\alpha$
is compensated by the change in sign of $m_\alpha$.
As a result the  {\em Chern number is unchanged
  from its initial value}, even if one quenches between different
phases.  Similar results may also be obtained for a linear sweep, $m_\alpha(t)=t/\tau$; see Supplemental Material.

\vssp \emph{Preservation of Chern Number.}--- An intuitive way to
understand the persistence of $\nu$ following a quantum quench is in
terms of spin-textures in momentum space. The Dirac Hamiltonian in
Eq.~(\ref{eq:dirac_matrix}) can be recast as an effective spin in a
${\bf k}$-dependent magnetic field, ${\bf h}_\alpha({\bf
  k})$. Explicitly, $\hat H_\alpha({\bf k})\equiv -{\bf h}_\alpha({\bf
  k})\cdot{\boldsymbol\sigma}/2$, where ${\boldsymbol\sigma}$ are the
Pauli matrices.
In equilibrium, the topological phases with $\nu_\alpha=\pm 1/2$
correspond to meron spin configurations which wind on the upper
(lower) half-sphere \cite{Fradkin}.  Following a quantum quench, the
spins precess in the effective magnetic field of the new Hamiltonian,
preserving the topological characteristics of the initial spin
configuration.  A similar argument may also be applied to the Haldane
model (\ref{eq:HM_realspace}) in ${\bf k}$-space. Indeed, one expects
the preservation of topological invariants under time evolution to be
a general feature for non-interacting fermions in a periodic system,
where each ${\bf k}$-state evolves unitarily under some Hamiltonian
$\hat{H}({\bf k})$, provided $\hat{H}({\bf k})$ is smoothly varying in
${\bf k}$-space. 

\begin{figure}[t] 
  \centering 
  \includegraphics{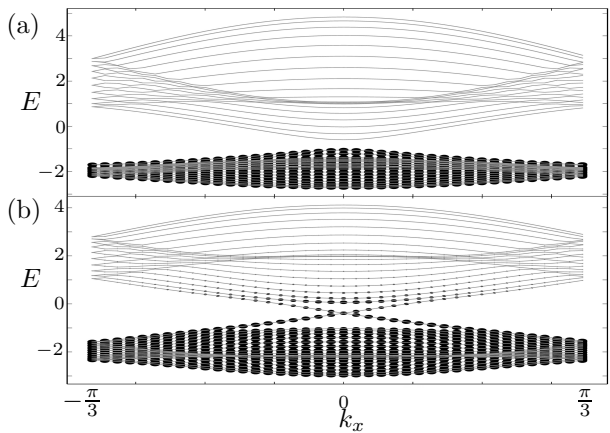}
     \caption{Energy spectrum of the Haldane model obtained by exact
       diagonalization on a finite-size strip of width $N=20$ unit
       cells with armchair edges. We take periodic (open) boundary
       conditions along (transverse to) the strip and set $t_1=1$,
       $t_2=\frac{1}{3}$, and $M=1$. (a) Equilibrium population of the
       energy levels in the non-topological phase with
       $\varphi=\frac{\pi}{6}$. (b) Re-population of the levels after
       a quench to the topological phase with
       $\varphi=\frac{\pi}{3}$, corresponding to the solid arrow in
       Fig.~\ref{fig:Hal_Ch} (a).  The size of the dots is
       proportional to the probability of finding a particle in the
       mode. Post-quench, the filling of the edge states and the bands
       is non-trivial.}\label{fig:occ}
\end{figure}

\vssp \emph{Edge States.}--- In the above discussion we have
demonstrated that the value of $\nu$ is unchanged as one quenches and
sweeps between different phases. However, there is a fundamental
distinction between the topological and non-topological phases, due to
the presence or absence of edge states in a finite-size sample
\cite{Hao2008}. In quenching between phases of different topological
character, these edge states will either appear or disappear,
depending on the direction of the quench. 
This is confirmed in Fig.~\ref{fig:occ}, which shows the
re-construction and re-population of the energy levels following a
quench from the non-topological phase to a topological phase. 
It is readily seen that the edge states emerge and are populated as a
result of the quench, in spite of the fact that $\nu$ remains equal to
zero in the absence of boundaries. Conversely, a quench from a
topological phase to the non-topological phase eliminates the
edge states, whilst $\nu$ remains pinned at unity.

\vssp \emph{Edge Currents and Orbital Magnetization.}--- Having
examined the re-population of the edge states we now consider physical
observables that depend on these states, including the edge currents
and the orbital magnetization. We first consider these quantities in
equilibrium, which already display interesting features. We define the
local current flowing through the site $i$ by
$\hat{{\bf J}}_i = - \frac{{\rm i}}{2}\sum_j {\boldsymbol
  \delta}_{ji}(t_{ij} \hat c^\dagger_i \hat c_j - {\rm h.c.})$, where
$t_{ij}$ is the hopping parameter of the Haldane model between sites
$i$ and $j$, ${\boldsymbol \delta}_{ji}$ is the vector displacement of
site $i$ from $j$, and the sum is over the nearest and next nearest
neighbors. The site indices may be decomposed into the triplet
$\{m,n,s\}$ labeling the $x$ and $y$ positions of the unit cell and
the sublattice index $s=A,B$. 
The total longitudinal current flowing along the strip in the $x$-direction at a definite
transverse $y$-position is therefore given by $J_n^x=\langle \hat{J}^x_n\rangle =\sum_{m
  s} \langle{\hat{J}^x_{mns}}\rangle$. 
\begin{figure}[ht]
  \centering 
  \includegraphics{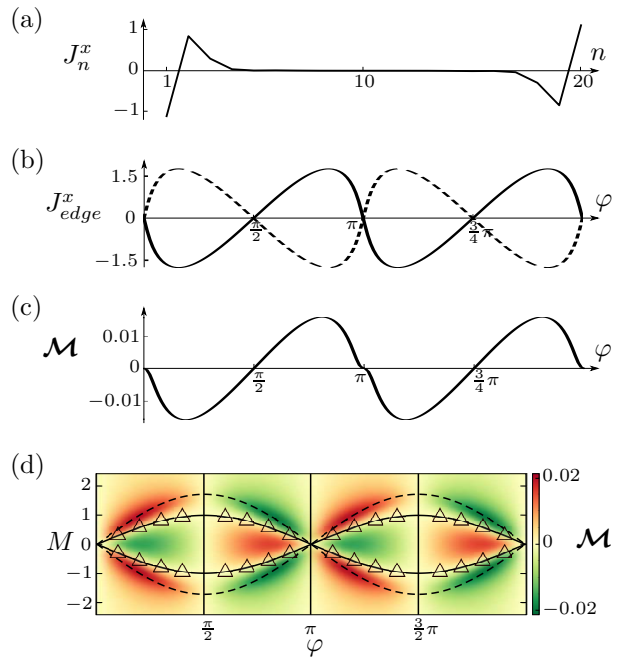}
     \caption{Equilibrium properties of the Haldane model on a
       finite-size strip as used in Fig.~\ref{fig:occ}.  (a) Total
       longitudinal current $J_n^x$ along the strip as
       a function of the transverse spatial index $n\in 1,\dots,20$,
       for $M=0$ and $\varphi=\pi/3$.  (b) Edge currents corresponding
       to $J_n^x$ with $n=1$ (solid) and $n=20$
       (dashed) for $M=0$. The edge currents exhibit $\pi$-periodicity
       in $\varphi$ and vanish when $\varphi=\pi/2$.  (c) Orbital
       magnetization ${\boldsymbol {\mathcal M}}$ as a function of
       $\varphi$ for $M=0$. (d) Intensity plot of ${\boldsymbol
         {\mathcal M}}$ where the dashed lines correspond to the
       boundaries of the topological phases. Numerically we observe
       that ${\boldsymbol {\mathcal M}}$ vanishes on the
       loci $M=\pm \sin\varphi$ (solid) within the topological
       phases. The loci are fits to the numerical data (triangles)
       where ${\boldsymbol {\mathcal M}}=0$. The magnetization also
       vanishes on the vertical lines $\varphi=\pi/2,\pi,3\pi/2,\dots$
       as follows from symmetry considerations.}
\label{fig:equil_edges}
\end{figure}
In Fig.~\ref{fig:equil_edges}(a) we plot
this current within the topological phase for $M=0$ and
$\varphi=\pi/3$. 
The presence of the counter-propagating edge currents is readily
seen. In Fig.~\ref{fig:equil_edges}(b) we show the dependence of these
edge currents on $\varphi$. Somewhat surprisingly, {\em the edge
  currents vanish within the topological phase, in spite of the
  presence of edge states in the spectrum}. The edge currents are
composed of counter-propagating contributions which cancel at
$\varphi=\pi/2$; see Supplemental Material. Moreover, the longitudinal
currents $J_n^x$ exhibit $\pi$-periodicity in $\varphi$. This is a
consequence of being at half-filling and occurs in spite of the fact
that the Hamiltonian and the current operator have a periodicity of
$2\pi$.  To prove the $\pi$-periodicity in $\varphi$ we first note
that both the Hamiltonian and the current operator change sign under the
transformation $M\rightarrow -M$, $\varphi \rightarrow \varphi+\pi$,
$\hat c_{mnA}\rightarrow\hat c_{mnA}$, $\hat c_{mnB}\rightarrow -\hat
c_{mnB}$, thereby interchanging the upper and the lower bands.  At
half-filling, we fill only the lower band, and it follows that
$J_n^x(M,\varphi)=J_n^x(-M,\varphi+\pi)$. In addition, the current
changes sign under the parity transformation $x\rightarrow -x$. This
interchanges the sublattices and corresponds to $M\rightarrow -M$ and
$\varphi\rightarrow -\varphi$. It follows that
$J_n^x(M,\varphi)=-J_n^x(-M,-\varphi)=J_n^x(-M,\varphi)$, where in the
last step we use the transformation properties under
time-reversal. Combining these relations, one obtains the
$\pi$-periodicity in $\varphi$ and the vanishing of the longitudinal
currents for $\varphi=\pi/2$.

Similar arguments also apply to the (lattice discretization of the)
orbital magnetization:
\begin{equation}
{\boldsymbol{\mathcal M}}=\frac{1}{2{\mathcal A}}\int d^2r\, {\bf
  r}\times \langle \hat {\bf J}({\bf r})\rangle,
\label{eq:orbital_magnetization}​
\end{equation}
where $\hat{\bf J}({\bf r})$ is the local current density operator and
${\mathcal A}$ is the area. As shown in Fig.~\ref{fig:equil_edges}(c)
this also vanishes within the topological phases and has
$\pi$-periodicity in $\varphi$
\cite{Thonhauser2005,Ceresoli2006}. 
Our numerical computations also reveal that the
magnetization vanishes on a sinusoidal locus $M=\pm \sin(\varphi)$
within the topological phases; see Fig.~\ref{fig:equil_edges}(d). In
addition, ${\boldsymbol {\mathcal M}}(M,\varphi)$ has extrema at $M=0$
and on the topological phase boundaries, $M=\pm \sqrt 3
\sin(\varphi)$, for fixed $\varphi$. Away from half-filling, the
particle-hole symmetry is broken and the periodicity of the currents
and the magnetization is restored to $2\pi$. The increase or decrease
of the edge currents depends on the sign of the doping and the Chern
index; see Supplemental Material.

\begin{figure}[ht]
  \centering 
  \includegraphics{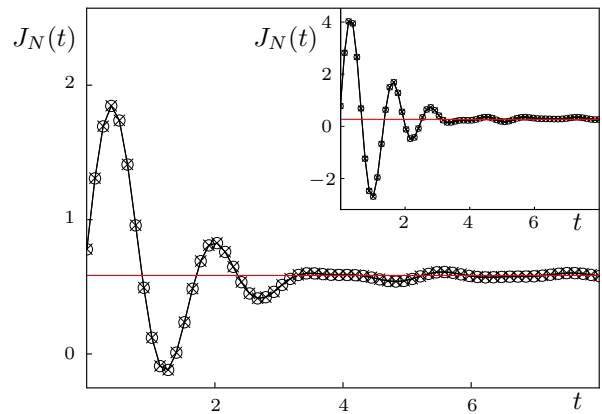}
     \caption{Dynamics of the edge current $J_N^x(t)$ for $N=30$
       (circles) and $N=40$ (crosses) following a quantum quench
       between the topological phase and the non-topological phase
       with $t_1=1$, $t_2=\frac{1}{3}$ and fixed
       $\varphi=\pi/3$. Quenches from $M=1.4$ to $M=1.6$ (main panel)
       and from $M=1.4$ to $M=2.2$ (inset) showing that the edge
       currents approach new equilibrium values.  For the chosen
       parameters, these are very close to the ground
       state expectation values of $J_N^x$ in the final Hamiltonian, as
       indicated by the horizontal lines.}
\label{fig:edge_dyn}
 \end{figure}

\begin{figure}[ht]
  \centering \includegraphics{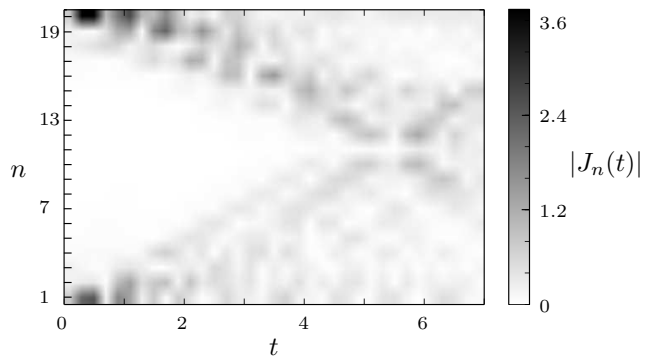}
     \caption{Dynamics of the  currents $\left|J_n^x(t)\right|$
       following a quantum quench from the topological to the
       non-topological phase for the parameters used in the inset of
       Fig.~\ref{fig:edge_dyn}. The damped oscillations of the edge
       currents are clearly visible, as is the light-cone spreading of
       the currents into the interior of the sample,
       where $c=3t_1/2\hbar=3/2$ is the effective speed of light. The
       waves propagating from the two edges meet at time $t\sim
       (N/2)\sqrt{3}/2c\sim 5.77$, leading to resurgent oscillations in
       finite-size samples; see Supplemental Information.}
\label{fig:jnt}
 \end{figure}

\vssp \emph{Dynamics of the Edge Currents.}--- Having discussed the
equilibrium properties of the edge currents we now consider their
response to quantum quenches. In Fig.~\ref{fig:edge_dyn} we show
quenches from the topological to the non-topological phase.  The edge
currents decay towards new values that are found to be numerically
close to the equilibrium values of the post-quench Hamiltonian. This
is in spite of the fact that the system is left in an excited state
under unitary evolution, and that $\nu$ remains pinned to unity in the
absence of boundaries. Quenches from the non-topological to
topological phases exhibit similar behavior; see Supplemental
Material.
Further insight into the non-equilibrium evolution may be gleaned from
the time-evolution of the longitudinal currents across the
two-dimensional system. As shown in Fig.~\ref{fig:jnt}, the damped
oscillations of the edge currents is accompanied by the light-cone
spreading of the currents into the interior of the sample.  It would
be interesting to observe this dynamics in experiment, which is in
principle possible if local imaging is available~\cite{Goldman2013a}.

\vssp \emph{Conclusions.}--- In this manuscript we have explored the
non-equilibrium dynamics of the Haldane model. We have demonstrated
that the Chern number is preserved in both quenches and sweeps between
different regions of the phase diagram. However, the edge states may
be re-constructed and re-populated leading to changes in the
accompanying edge currents. Predictions for experiment include the
vanishing of the equilibrium edge currents in the topological phases,
and the light-cone spreading of the currents following a quantum
quench. There are a wide variety of directions for further research,
including the dynamics of the conductivity and its relation to the
Chern number, and the role of decoherence via coupling to the
environment.

\emph{Acknowledgements.}--- This work was supported by EPSRC Grants
EP/J017639/1 and EP/K030094/1. MJB thanks the EPSRC Centre for
Cross-Disciplinary Approaches to Non-Equilibrium Systems (CANES)
funded under grant EP/L015854/1. MJB and MDC thank the Thomas Young
Center.
\vssp

Whilst this work was in preparation the pre-print \cite{DAlessio2014}
appeared, which reaches similar conclusions to ours regarding the
invariance of $\nu$ under unitary evolution.

\onecolumngrid

\vspace{0.5cm}

\section{Supplemental Material}

\twocolumngrid

\appendix

\emph{Haldane Model.}--- For completeness, let us recall some details
of the Haldane model~\cite{Haldane1988}. A key feature is the presence
of time-reversal symmetry breaking induced by the complex
second-neighbor hopping. In Haldane's original paper this corresponds
to a staggered magnetic field, where the local flux within a hexagonal
plaquette is non-zero, but the total flux through a plaquette
vanishes. In the experimental realization of Ref.~\cite{Esslinger2014}
this is achieved by circular modulation of the lattice position.  In
Fourier space, the Hamiltonian is given by~\cite{Haldane1988}
\begin{multline}
H(\mathbf{k})=2 t_2 \cos \varphi \left(\sum_i \cos(\v k \cdot \v b_i)\right) {\rm I} \\
+ t_1\left(\sum_i [\cos(\v k\cdot \v a_i)\sigma^1 +\sin(\v k \cdot \v a_i)\sigma^2]\right)\\
+ \left[M-2t_2\sin\varphi \left(\sum_i \sin(\v k \cdot \v b_i)\right)\right] \sigma^3,
\end{multline}
where $\sigma^\mu$ are Pauli matrices and ${\rm I}$ is the identity
matrix.  Here, $\v a_1,\v a_2,\v a_3$ are the displacements from a $B$
site to its three nearest-neighbor $A$ sites. The $\v b_i$'s are
defined via cyclic permutations of $\v b_1:=\v a_2 -\v a_3$ and
correspond to the second-neighbor displacements, $\pm \v b_i$. For
$M=0$ and $\varphi=0$, the two bands touch at the six corners of the
Brillouin Zone. However, only two of these points are inequivalent, 
and we denote these by $\v k_\alpha$ with $\alpha=\pm$. Expanding the
Hamiltonian around these two corners, 
with $\vg\Pi_\alpha=(\Pi^x_\alpha,\Pi^y_\alpha)=\v k-\v k_\alpha$, one
obtains~\cite{Haldane1988}
\begin{equation}
 H_\alpha= c(\Pi^1_\alpha \sigma^2-\Pi^2_\alpha \sigma^1) + m_\alpha c^2 \sigma^3,
\end{equation}
where $m_\alpha$ and $c$ defined following Eq.~(\ref{eq:dirac_matrix}) and
\begin{equation}
 (\Pi^1_\alpha+\iu \Pi^2_\alpha)=\frac{2}{3} \sum_i e^{\iu \v k_\alpha\cdot \v a_i}\frac{\v a_i \cdot \vg \Pi_\alpha}{\abs{\v a_i}}.
\end{equation}
In order to connect with the notations used in Eq.~(\ref{eq:dirac_matrix})
we parameterize the complex number $\Pi^2_\alpha+\iu
\Pi^1_\alpha$ as $k e^{\iu\alpha\theta}$.

\vssp \emph{Berry Phase.}--- In a gauge theory, 
invariant under a local transformation $\Ket{\psi\p(\v
  X)}:=e^{-\iu \rho(\v X)} \Ket{\psi(\v X)}$, parameterized by $\v
X$, one may define a covariant derivative
\begin{equation}
 \Ket{D_\mu\psi(\v X)}=\Ket{\partial_\mu \psi(\v X)}-\Ket{\psi(\v X)}\Braket{\psi(\v X)|\partial_\mu \psi(\v X)},
\end{equation}
where $\Ket{\psi(\v X)}\Bra{\psi(\v X)}$ projects out the parts of
$\Ket{\partial_\mu \psi(\v X)}$ that are not orthogonal to
$\Ket{\psi(\v X)}$. One may also introduce the Berry connection $
A_\mu(\v X)=\iu \Braket{\psi(\v X)|\partial_\mu \psi(\v X)} $, the
Berry phase $\phi$, for a closed path $\Gamma$ in parameter space $
e^{\iu \phi}:=e^{\iu \oint_\Gamma \diff X^\mu A_\mu(\v X)}$, and the
Berry curvature $\Omega_{\mu\sigma}= \Braket{D_\mu\psi(\v X)|D_\sigma
  \psi(\v X)}-\Braket{D_\sigma\psi(\v X)|D_\mu \psi(\v X)}$. One may
also generalize the Gauss--Bonnet theorem:
\begin{equation}
 \nu := \frac{\phi}{2\pi}= \frac{1}{2\pi}\int_{M_2} \diff X^\mu \wedge \diff X^\sigma\,\Omega_{\mu \sigma}(\v X) \in {\mathbb Z},
\end{equation}
where $M_2$ is a closed orientable 2-manifold and $\nu$ is the
Chern number. 

\vssp 

\emph{Chern Number for Dirac Points.}--- For the Dirac Hamiltonian,
given by Eq.~(\ref{eq:dirac_matrix}) with $\alpha=-1$ say,
the Berry connection is given by $A_k^- = \iu \Braket{\psi|\partial_k \psi}$ and $A_\theta^-=\iu \Braket{\psi|\partial_\theta \psi}$, where we parameterize
the momentum-space 2-manifold $M_2$ by $(k,\theta)$.
For a superposition of the form
\begin{equation}
\Ket{\psi(k,\theta)}=a(k)\Ket{l(k,\theta)}+b(k)\Ket{u(k,\theta)},
\end{equation}
where $|l\rangle$ and $|u\rangle$ correspond to the lower and upper
band eigenstates, 
$A_k^-$ and $A_\theta^-$ are independent of $\theta$.  As a result,
the contribution of this Dirac point to the Chern number,
$\nu=\nu_++\nu_-$, is
\begin{equation}
 \nu_-=\frac{1}{2\pi}\int_0^{2\pi} \diff \theta \int_0^\infty \diff
 k\,\Omega^-=\int_0^\infty \diff k\,\partial_k A_\theta^- =
 \left. A_\theta^- \right|_0^\infty,
\label{eq:numinus}
\end{equation}
where $\Omega^-=\partial_kA^-_\theta-\partial_\theta A_k^-$.
Using the explicit forms 
\begin{equation}
\Ket{l}   = \begin{pmatrix}
                                 e^{-\iu \theta} f_-(k,m_-)\\ f_+(k,m_-)
                                \end{pmatrix},\quad 
\Ket{u} = \begin{pmatrix}
                                 -e^{-\iu \theta} f_+(k,m_-)\\ f_-(k,m_-)
                                \end{pmatrix}
\end{equation}
where $f_\pm(k,m):=\sqrt{\frac{1}{2}\left(1\pm\frac{m_-}{\sqrt{k^2+m_-^2}}\right)}$,
one obtains
\begin{multline}
 A_\theta^-=\abs{a(k)}^2 f_-^2(k,m_-)+\abs{b(k)}^2 f_+^2(k,m_-)\\-\frac{k}{2\sqrt{k^2+m_-^2}}\left(a\conj(k)b(k)
 + {\rm h.c.}\right).\label{eq:berry_conn}
\end{multline}
It follows that
\begin{align}
 A_\theta^-(\infty)   &    =  \frac{1}{2}\left[1-\left(a\conj(\infty)b(\infty) + {\rm h.c.}\right)\right] \\
 A_\theta^-(0)           &    =  \begin{cases}
                               \abs{b(0)}^2 & m_->0 \\ \abs{a(0)}^2 & m_-<0.
                               \end{cases}
\end{align}
Eq.~(\ref{eq:chern_number}) follows by reinstating
the time dependence $a(k)\rightarrow a(k)e^{-\iu E_l t}$ and
$b(k)\rightarrow b(k)e^{-\iu E_u t}$, for unitary evolution under the
Dirac Hamiltonian.

\vssp
{\em Quench.}--- We now consider a quench $m_- \rightarrow m_-\p$. 
The system is initially prepared in the ground state of $H_-(m_-)$,
corresponding to the filled lower band, $\Ket{l}$. Immediately after
the quench we may decompose $\Ket{l}$ into the eigenstates of the 
post-quench Hamiltonian:
\begin{align*}
a(k) & = f_-(k,m_-)f_-(k,m_-\p)+f_+(k,m_-)f_+(k,m_-\p), \\
b(k) & = f_+(k,m_-)f_-(k,m_-\p)-f_-(k,m_-)f_+(k,m_-\p).
\end{align*}
The probability of being in the upper band, $|b_-(k)|^2$ is plotted in
Fig.~\ref{fig:bsquare}. In particular, one obtains $b(\infty)=0$ and $b(0)=0$ ($b(0)=\pm 1$) for sign-preserving (sign-changing) mass quenches.  
The preservation of the contribution to the 
Chern number is discussed in the main text. 

\vssp \emph{Linear Sweep.}--- We extend the results for the Chern
number preservation to linear, time-dependent, sweeps. We again
consider $\alpha=-$ and set $m_-=\frac{t}{\tau}$ in
Eq.~(\ref{eq:dirac_matrix}), corresponding to a sign changing sweep over the interval $t\in [-\infty,\infty]$. We prepare the system in the ground state
at $t=-\infty$ and track the subsequent evolution \cite{Dutta2010}. At
any instant of time the state can be written as a superposition over
the eigenstates of $H_-(t)$:
\begin{equation} 
\Ket{\psi(k,\theta,t)}=a(k,t)\Ket{l(k,\theta,t)}+b(k,t)\Ket{u(k,\theta,t)}.
\end{equation}
Using Eq.~(\ref{eq:numinus}) and Eq.~(\ref{eq:berry_conn}) one obtains
\begin{multline}
\nu_-(t)=\sign t \left(\frac{1}{2}-\abs{b(0,t)}^2\right) \\ + \left[\lim_{k\rightarrow \infty} -\lim_{k\rightarrow 0}\right] \left[\frac{k}{2\sqrt{k^2+\frac{t^2}{\tau^2}}}(a\conj(k,t)b(k,t) + \mbox{h.c.})\right].\label{eq:chernsweep}
\end{multline}
We now need to evaluate the coefficients $a$ and $b$ for $k=0$ and $k=\infty$. For $k=0$ the eigenstates of $H_-(t)$ are
\begin{align}
\Ket{l(0,t)} &= \begin{pmatrix}
			e^{-\iu \theta} \sqrt{\frac{1}{2}(1-\sign{t})} \\ 
			\sqrt{\frac{1}{2}(1+\sign{t})}
		  \end{pmatrix}, \\
\Ket{u(0,t)} &= \begin{pmatrix}
			-e^{-\iu \theta} \sqrt{\frac{1}{2}(1+\sign{t})} \\
			\sqrt{\frac{1}{2}(1-\sign{t})}
		  \end{pmatrix}.
\end{align}
These are time-independent for $t\neq 0$, and for $k=0$ the system
remains in the lower band up to $t=0^-$.  At $t=0^+$ the mass
parameter changes sign, and at $k=0$, overlaps completely 
with the upper band:
$$\abs{\Braket{l(0,0^-)|u(0,0^+)}}=1.$$ Henceforth, the $k=0$ mode
remains in 
the upper band.  At $k=\infty$ the eigenstates are 
independent of time:
\begin{align}
\Ket{l(\infty,t)} &= \begin{pmatrix}
			e^{-\iu \theta} \sqrt{\frac{1}{2}} \\ 
			\sqrt{\frac{1}{2}}
		  \end{pmatrix}, \\
\Ket{u(\infty,t)} &= \begin{pmatrix}
			-e^{-\iu \theta} \sqrt{\frac{1}{2}} \\
			\sqrt{\frac{1}{2}}
		  \end{pmatrix}.
\end{align}
The system remains in the lower band for $k=\infty$.  Summarizing, one obtains
\begin{equation}
 \abs{b(0,t)} =\begin{cases}
         0, & t<0 \\ 1, & t>0
        \end{cases} \quad {\rm and} \quad 
 \abs{b(\infty,t)} = 0\quad \forall t. \label{eq:sweep2}
\end{equation}
This parallels the situation for the quench protocol as shown by the
dashed line in Fig.~\ref{fig:bsquare}.  The preservation of $\nu_-$ 
follows by substituting Eq.~(\ref{eq:sweep2}) into
Eq.~(\ref{eq:chernsweep}).

\begin{figure}[h]
  \centering
   \includegraphics{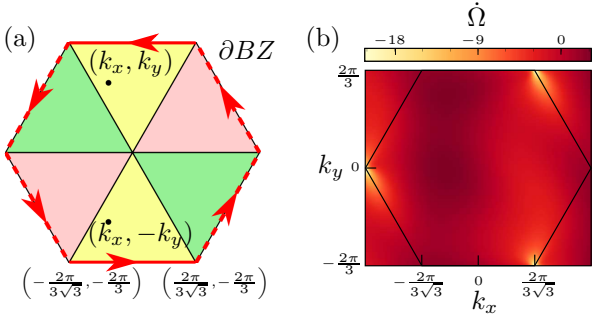}
     \caption{First Brillouin zone of the Haldane model. (a) In each
       of the triangles the Berry connection and the Berry curvature
       are time-dependent. However, the Chern number is given by the
       line integral of the Berry connection along the zone boundary,
       as indicated by the arrows.  Following a quench, the
       time-dependent contributions to $\nu$ from opposite sides of
       the boundary cancel each other. (b) Time derivative of the
       Berry curvature for $M=1$, $t_1=1$ and $t_2=1/3$ following
       a quench from the non-topological phase with $\varphi=\pi/6$
       to the topological phase with $\varphi=\pi/3$. Although $\dot\Omega\neq 0$,
       numerical integration over the Brillouin zone confirms that
       $\dot\nu=0$.}\label{fig:BZsym}
 \end{figure}

\vssp \emph{Preservation of Chern Number.}--- Having established the
preservation of $\nu$ using the low-energy Dirac Hamiltonian, we
examine the non-equilibrium response of the Haldane model. By
recasting Eq.~(\ref{eq:chernomega}) in the form
\begin{equation}
\nu = \frac{1}{2\pi}\oint_{\partial {\rm BZ}} \diff k^\mu A_{k_\mu}, \label{eq:bp}
\end{equation}
we may focus on the time-dependence of the Berry connection.  In
general, $\dot A_{k_\mu}=i\langle
\dot\psi|\partial_{k_\mu}|\psi\rangle+i\langle\psi|\partial_{k_\mu}|\dot\psi\rangle=\langle
\psi|[\partial_{k_\mu},\hat H]|\psi\rangle$, or equivalently, $\dot
A_{k_\mu}=\langle \psi|(\partial_{k_\mu}\hat H)|\psi\rangle$.  Expanding 
the initial state $|\psi\rangle=\sum_{\gamma=l,u}c_\gamma e^{-iE_\gamma t}|\gamma\rangle $ in terms of the eigenstates $|\gamma\rangle$ of the final Hamiltonian
\begin{equation}
\dot A_{k_\mu}(k_x,k_y)=\sum_{\gamma=l,u}c_\gamma
c_{\gamma^\prime}^\ast \langle \gamma^\prime|\partial_{k_\mu}\hat
H|\gamma\rangle e^{i(E_{\gamma^\prime}-E_\gamma)t}.
\end{equation}
In general, this is time-dependent. However, using the symmetries of
the final Hamiltonian, the components of $\dot A_{k_\mu}$ along the
Brillouin zone boundary occur in equal pairs and cancel in the line
integral for $\dot \nu$.  For example, within the upper and lower
triangles depicted in Fig.~\ref{fig:BZsym}, $ \hat H(k_x,k_y)=\hat 
H\conj(k_x,-k_y)$. Periodicity in $k$-space 
ensures that  $\hat H(k_x,k_y)=\hat H(k_x,-k_y)$ 
along the corresponding zone boundaries, so that 
$\hat H$ is real on these segments. 
It follows that $\dot A_{k_x} (k_x,2\pi/3)=\dot A_{k_x} (k_x,-2\pi/3)$
for $k_x \in (-\frac{2\pi}{3\sqrt{3}},\frac{2\pi}{3\sqrt{3}})$, so
these two contributions to $\dot\nu$ cancel.

\begin{figure}
  \centering \includegraphics{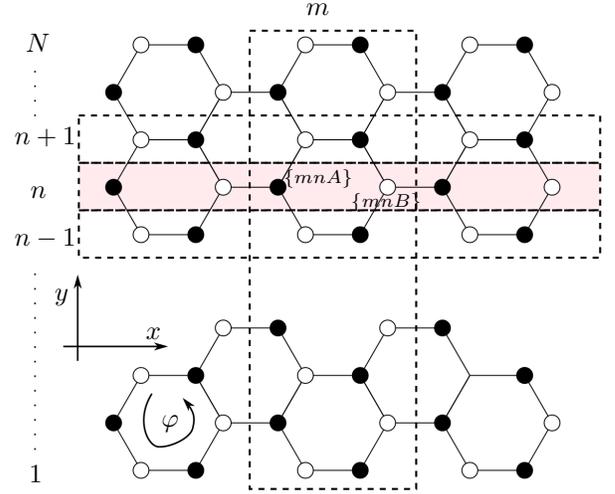} \caption{The
   strip geometry used for our finite-size computations.
   The strip is $N$ cells wide with arm chair edges and has periodic
   (open) boundary conditions in the longitudinal (transverse)
   directions. Each unit cell is labeled by two indices $m$ and $n$,
   and contains two sites belonging to the $A$ or $B$ sublattices. The
   phase $\varphi$ of the Haldane model is taken as positive for
   anticlockwise next to nearest neighbor
   hopping.}\label{fig:lattice} \end{figure}

\begin{figure}
  \centering \includegraphics{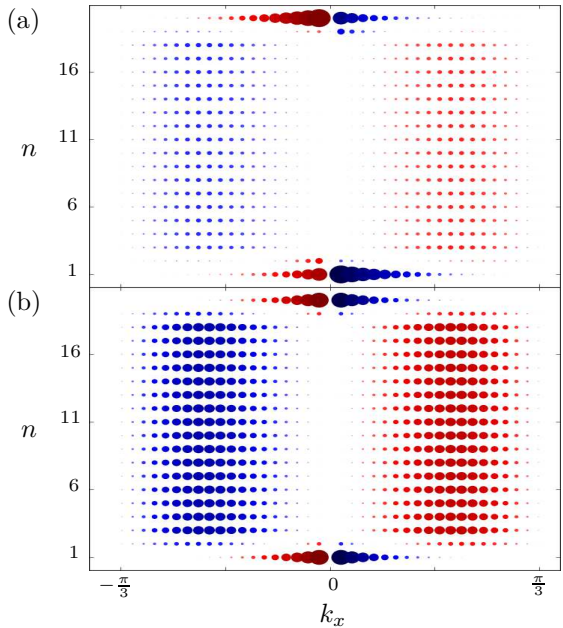} 
\caption{Momentum 
space contributions to the equilibrium currents along the strip,
   $\langle \hat J_n^x(k_x)\rangle$, with $N=20$, $t_1=1$, $t_2=1/3$
   and $M=0$.  For clarity, the size of the dots is proportional to
   the fourth power of $\langle \hat J_n^x(k_x)\rangle$ and the blue
   (red) dots indicate negative (positive) values. (a) $\varphi=\pi/3$
   showing counter-propagating contributions to the 
   currents. The net currents vanish in the bulk but are non-zero
   close to the edges; see Fig.~\ref{fig:equil_edges}(a).  (b)
   $\varphi=\pi/2$ showing balanced contributions throughout the
   strip, leading to $\langle \hat J_n^x\rangle =0$.
   }\label{fig:counterProp} \end{figure}

\begin{figure}
  \centering \includegraphics{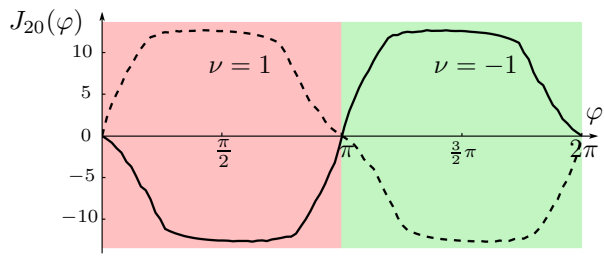} 
\caption{Equilibrium
     edge current $J_N^x$ for $N=20$, $t_1=1$, $t_2=1/3$ and $M=0$
     with a $5\%$ particle (solid) or hole (dashed) doping.  In
     contrast to the half-filled case shown in
     Fig.~\ref{fig:equil_edges}(b), the currents no longer vanish at
     $\varphi=\pi/2$. Instead, the currents have the same periodicity
     as the Hamiltonian. The increase or decrease of the edge current
     at $\varphi=\pi/2$ reflects both the sign of the doping and the
     Chern index. }\label{fig:J20doped} \end{figure}

\vssp \emph{Edge Currents and Orbital Magnetization.}---
In order to examine the behavior of the edge currents we consider the
Haldane model on a finite-size strip with armchair edges and periodic
(open) boundary conditions along (transverse to) the strip. The
geometry we use is shown in Fig.~\ref{fig:lattice}. The numerical
computations in the main text are performed on strips of width $N=20$,
$30$ or $40$ unit cells.
We define the longitudinal current $\hat J_n^x$ as the total current
flowing along the $n$-th row of the strip where $n\in 1,\dots,N$, as
shown by the shaded area in Fig.~\ref{fig:lattice}. Exploiting the
periodic boundary conditions along the strip, $\hat
J_n^x=\sum_{k_x}\hat J_n(k_x)$. As shown in
Fig.~\ref{fig:counterProp}, the terms in this summation appear with
both positive and negative signs. In particular, the edge currents,
$\langle \hat J_1^x\rangle$ and $\langle \hat J_{N}^x\rangle$, receive
opposite contributions which perfectly cancel when $\varphi=\pi/2$.
Doping the system with particles or holes breaks this symmetry and
restores the $2\pi$-periodicity, as shown in
Fig.~\ref{fig:J20doped}. 
The effect of doping is to change the
relative contributions of the bulk and the edge states to the total
edge currents; for particle (hole) doping the edge (bulk) states
contribute more to the overall edge current.

\begin{figure}
  \centering \includegraphics{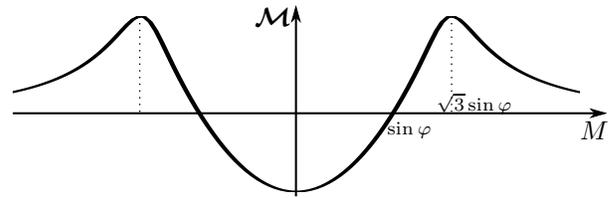} \caption{Orbital
   magnetization $\boldsymbol{\mathcal M}$ with $N=20$, $t_1=1$,
   $t_2=1/3$ and $\varphi=\pi/3$. Numerically we observe that
   $\boldsymbol{\mathcal M}$ vanishes within the topological phases
   when $M=\pm\sin \varphi$. We also observe that
   $\boldsymbol{\mathcal M}$ has extrema for $M=0$ and
   $M=\pm\sqrt{3}\sin \varphi$; the latter correspond to the
   boundaries of the topological phases as indicated by the dotted
   lines.}\label{fig:mag_M} \end{figure}

Zeros of the orbital magnetization also occur within the topological
phases. As shown in Fig.~\ref{fig:mag_M}, ${\boldsymbol{\mathcal
M}}(M,\varphi)$ has extrema at $M=0$ and on the boundaries of the
topological phases at $M=\pm \sqrt{3}\sin\varphi$. Numerically we
observe that the zeros of ${\boldsymbol{\mathcal M}}$ occur on the
sinusoidal loci $M=\pm\sin \varphi$.
Once again, the effect of doping is to change the relative
contributions of the bulk and the edge states to the orbital
magnetization; for particle (hole) doping the edge (bulk) states
contribute more to $\boldsymbol{\mathcal M}$.

\begin{figure}
  \centering \includegraphics{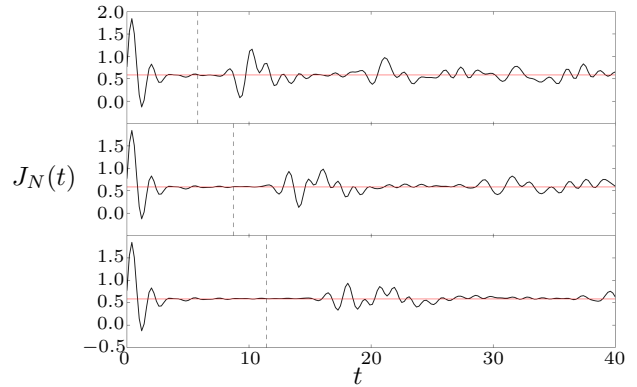} 
\caption{Time
     dependence of the edge current $J_N^x(t)$ after a quantum quench
     from the topological phase with $M=1.4$ to the non-topological
     phase with $M=1.6$, keeping $t_1=1$, $t_2=1/3$ and
     $\varphi=\pi/3$ fixed. This corresponds to the main panel of
     Fig.~\ref{fig:edge_dyn} over a longer time duration. The three
     panels from the top to the bottom correspond to strips of width
     $N=20$, $30$ and $40$ respectively. The appearance of resurgent
     oscillations is evident in all three panels due to the finite
     width of our system. The dashed lines indicate the time-scale
     $t=(N/2)\sqrt{3}/2c$ at which signals propagating from the two
     edges meet, corresponding to the onset of finite-size
     effects. The horizontal line corresponds to the ground-state
     expectation value of the edge current for the post quench
     Hamiltonian.}
\label{fig:long_time_th_limit}
\end{figure}

\vssp \emph{Dynamics of the Edge Currents.}--- Following a quantum
quench, we observe that the edge currents relax to new values that are
comparable to the ground expectation values evaluated for the final
Hamiltonian. In these examples we necessarily see finite-size effects
due to the finite width of the strip. At late times, we see resurgent
oscillations due to the light-cone propagation of currents into the
interior of the sample; see Fig.~\ref{fig:long_time_th_limit}.
The onset timescale for these resurgent oscillations increases with
the width of the strip. This timescale is given by $t=d/2c$ where
$d=N\sqrt{3}/2$ is the width of the sample and $c=3t_1/2\hbar$ is the
effective speed of light. This timescale is indicated by the dashed
lines in Fig.~\ref{fig:long_time_th_limit}. In order to avoid
finite-size effects in our predictions we therefore restrict the
domain of our simulations to be within this time interval. The
  agreement between the results for $N=30$ and $N=40$ in
  Fig.~\ref{fig:edge_dyn} of the main text highlights that we are
  probing the intrinsic dynamics of the edge currents, before
  finite-size effects play a role.

\begin{figure}[p]
  \centering \includegraphics{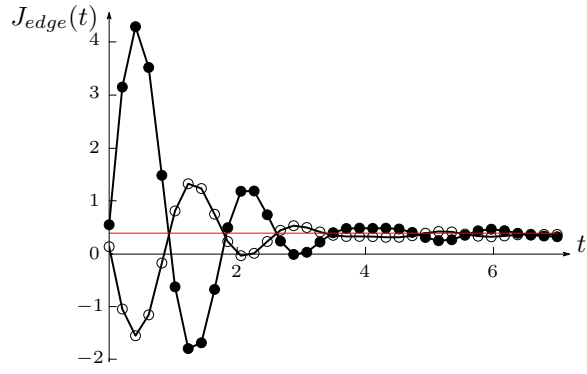} \caption{Edge current $J_N^x(t)$
      following a quantum quench within the topological phase
      (filled circles) and from the non-topological phase to the
      topological phase (empty circles). We set $N=30$, $t_1=1$,
      $t_2=1/3$ and $\varphi=\pi/3$ and consider quenches of the mass
      parameter $M=0.5 \rightarrow 1.4$ (full circles) and
      $M=2.2\rightarrow1.4$ (empty circles).  The horizontal line
      corresponds to the ground-state expectation value of the edge
      current for the post quench Hamiltonian. The coincidence
      between the oscillation frequencies is consistent with the fact
      that we quench to the same final
      Hamiltonian.}\label{fig:Jtt_Jntt} \end{figure}

For completeness, in Fig.~\ref{fig:Jtt_Jntt} we show quenches
to the topological phase. Numerically we observe that the edge
currents approach new values that are very close to those evaluated in
the ground state of the final Hamiltonian. Likewise, the oscillation
frequencies coincide for quenches to the same final Hamiltonian.
\end{document}